\documentclass[namedreferences]{SolarPhysics}
\usepackage[optionalrh]{spr-sola-addons} 
\usepackage{graphicx}        
\usepackage{color}           
\usepackage{url}             

\usepackage{solaheader}	
\newcommand{\solareceiveddate}{9 July 2012} 
\newcommand{\solaaccepteddate}{21 January 2013}
\newcommand{\solaonlinedate}{8 February 2013}
\newcommand{\doi}{10.1007/s11207-013-0241-8}
\renewcommand{\solayear}{2013}

\usepackage{hyperref}  
\definecolor{lightblue}{rgb}{0.0,0.5,0.9}
\hypersetup{colorlinks={true}, urlcolor=lightblue, citecolor=lightblue}                                                                                                                  
\ifx \doiurl    \undefined \def \doiurl#1{\href{http://dx.doi.org/#1}{\textsf{#1}}}\fi
\ifx \adsurl    \undefined \def \adsurl#1{\href{http://adsabs.harvard.edu/abs/#1}{\textsf{#1}}}\fi
\ifx \arxivurl  \undefined \def \arxivurl#1{\href{http://arxiv.org/abs/#1}{\textsf{#1}}}\fi

\begin{document}
\begin{article}
\begin{opening}

\title{Magnetic Flux Change and Cancellation Associated with X-class and M-class Flares}

\author{Olga Burtseva and Gordon Petrie}

\institute{National Solar Observatory, 950 North Cherry Ave.,Tucson, AZ 85719, USA
 (email: \href{burtseva@noao.edu}{burtseva@noao.edu})}

\begin{abstract}

We perform a statistical study of permanent changes in longitudinal fields associated with solar flares by tracking magnetic features. The YAFTA feature tracking algorithm is applied to GONG++ one-minute magnetograms for 77 X-class and M-class flares to analyze the evolution and interaction of the magnetic features and to estimate the amount of canceled magnetic flux. We find that significantly more magnetic flux decreases than increases occurred during the flares, consistent with a model of collapsing loop structure for flares. Correlations between both total (unsigned) and net (signed) flux changes and the GOES peak X-ray flux are dominated by X-class flares at limb locations. The flux changes were accompanied in most cases by significant cancellation, most of which occurred during the flares. We find that the field strength and complexity near the polarity inversion line are approximately equally important in the flux cancellation processes that accompany the flares. We do not find a correlation between the flux cancellation events and the stepwise changes in the magnetic flux in the region. 

\end{abstract} 


\end{opening}

\section{Introduction}
It is now known that photospheric magnetic fields change significantly, abruptly and permanently as a result of major X- and M-class flares ({\it e.g.} \opencite{Cameron99}; \opencite{Kosovichev01}; \opencite{Wang02}, \citeyear{Wang12};  \opencite{Sudol05}; \opencite{Wang10}; \opencite{Petrie10}; \opencite{Gosain12}; \opencite{Sun12}; \opencite{Petrie12a}). Detection and study of photospheric magnetic field changes during solar flares can aid in a better understanding of the mechanisms of energy buildup and release in flares. Despite these observations, most models of flares assume that these events are coronal and that no photospheric magnetic field changes occur over the flare duration \cite{Priest02}. Ideally, coronal field models should be based on boundary conditions observed in the lower atmosphere. However, it is not clear how to relate the restructuring of approximately force-free or current-free coronal field with the changes in the forced photosphere. 

A few possible mechanisms were offered in past works that could cause the sudden and permanent photospheric field changes observed. The most frequently mentioned processes are the emergence of new magnetic flux and changes in the direction of the field \cite{Spirock02,Sudol05}. Supporting magnetic flux emergence as a possible mechanism, \inlinecite{Wang93a} observed emergence of new flux in sunspot umbra immediately after an X-class flare. \inlinecite{Fletcher08} suggested the Alfv\'en wave scenario as an explanation for the observed rapid variations in the line-of-sight component of the photospheric magnetic field in the flare impulsive phase: Large-scale Alfv\'en wave pulses may transport both energy and changes in the magnetic field rapidly from the flare site in the corona to the lower atmosphere. \inlinecite{Hudson08} estimated back reaction of the photospheric magnetic field during a  solar flare and proposed a model of magnetic implosion and collapsing loop structure towards the neutral line. \inlinecite{Wang10}, \inlinecite{Petrie10},\inlinecite{Wang12}, and \inlinecite{Petrie12} found strong evidence that this prediction is true. \inlinecite{Petrie10} analyzed pixel-by-pixel the field changes accompanying 77 X- and M-class flares using Global Oscillation Network Group (GONG) full-disk magnetograms. In this work, we use the same flares from the GONG flare list to study flare-related photospheric field changes by decomposing the field into individual magnetic features.  

As \inlinecite{Sudol05} have noted, the change in magnetic flux may be a more important physical quantity than the largest local field change at one particular location. By only picking up representative pixels with the fastest and largest field changes, free of artifacts \cite{Sudol05, Petrie10}, one may not capture a complete picture of the changes in the magnetic field in the region. However, the value of magnetic flux changes may capture the effects of flares on the entire active region. The calculations of magnetic flux made by past authors have either summed the flux of changing pixels over the entire active region \cite{Petrie10} or confined attention to simple bipolar active regions with highly sheared neutral lines ({\it e.g.} \opencite{Wang06}; \opencite{Wang10}). In the present work the active regions are divided into magnetic features (substructures) whose flux changes are calculated individually. This allows us to identify which parts of a region change significantly and also to investigate the interactions among different features in an active region.

One of the processes involving magnetic features associated with flares is flux cancellation. Flux cancellation plays an important role in some theories of solar eruptions ({\it e.g.} \opencite{vanBallegooijen89}; \opencite{Amari10}). It is found to be essential for the formation of prominences \cite{Martin98}, and may possibly trigger filament eruptions and coronal mass ejections \cite{Wang96,Zhang01}. The mechanism for flux cancellation is suggested to be a necessary condition of flare initiation as a part of slow reconnection processes in the lower atmosphere \cite{Martin85,Livi89,Mathew00}. The first evidence of the possibility of magnetic reconnection in flux cancellation was presented by \inlinecite{Wang93}, who analyzed vector magnetic field changes and demonstrated how features previously not directly connected by transverse fields partially reconnected during flux cancellation. Strong observational evidence of magnetic reconnection in the lower atmosphere leading to instability and flare/coronal mass ejection (CME) events was presented by \inlinecite{Zhang01}, \inlinecite{Kim01}, and \inlinecite{Moon04}. 

Reconnection can occur at different levels of the solar atmosphere; above, below, or at the photospheric level \cite{Zwaan87}. \inlinecite{Harvey99} observed that magnetic flux disappeared in the chromosphere earlier than in the photosphere at the cancellation sites, which suggests submergence of the flux implying that reconnection occurs above the photosphere. \inlinecite{Chae04} confirms this finding. However, \inlinecite{Yurchyshyn01} reported flux cancellation, following a C-class flare, with the presence of upflows in the canceling feature, which suggests the reconnection below or at the level of the photosphere. In the case of the formation of prominences, cancellation is explained by the emergence of U-shaped loops \cite{Lites95,Low01,Welsch05}. Most of these studies suggest flux cancellation as a pre-flare condition, with some exceptions ({\it e.g.} \inlinecite{Yurchyshyn01}). Photospheric magnetic shear increases have been observed to occur during and after flares \cite{Wang92}, and before and during flares \cite{Wang96a}. In this work, we estimate flux cancellation in the magnetic features at the polarity inversion line (PIL) associated with a major flare, at the time before, during, and after the flare.

\section{Data Analysis}
In this study full-disk photospheric magnetograms from the GONG sites are used. GONG full-disk magnetograms are obtained with a 1-minute cadence, 2.5-arcsec pixel size, and a noise level of 3 gauss (G) per pixel. We have analyzed 77 sets of GONG magnetograms for 38 X- and 39 M-class flares. We confined our attention to the most energetic flares with the best data coverage. We eliminated all flares of Geostationary Operational Environmental Satellite (GOES) class weaker than M5.0 and all with an apparent central meridian distance (longitude distance from central meridian) greater than $65^\circ$. We further limited our attention to those flares for which GONG magnetograms are available over approximately 1 hour before and after the flare. The flares studied are identified in Tables 1 (M-class) and 2 (X-class) of \inlinecite{Petrie10}. \inlinecite{Petrie10} studied, pixel by pixel, the field changes accompanying the flares. In this work we study flare-related photospheric field changes analyzing collections of pixels (magnetic features). 

The preliminary data reduction procedure is as in \inlinecite{Petrie10}. \textit{i}) The magnetograms are remapped to azimuthal equidistant projection, $32^\circ \times 32^\circ$ field of view, local to a flare. \textit{ii}) The maps are corrected for differential rotation and registered, as a reference, to  a ten-minute average of magnetograms prior to the flare.

We use the Yet Another Feature Tracking Algorithm (YAFTA) feature-tracking code \cite{Welsch03} to decompose the field into individual magnetic features and to study their evolution and interplay during the flare. We have used YAFTA's ``downhill'' algorithm to group together more than four pixels lying on the same ``hill'' in the absolute field strength above 100 G. The 100 G threshold identifies active region fields in GONG magnetograms. Small features often have short lifetimes. The smallest feature that showed abrupt and permanent flux change in this work consisted of eight pixels (which is $\approx$ 0.5 heliographic degrees in diameter), and the biggest one was 650 pixels ($\approx$ 3 degrees in diameter). The program accomplishes matching of features between two consecutive magnetograms. Most of the features fragment, merge, appear, and disappear between the magnetograms, which makes the matching process complex. As a result, features change their labels. Only features that do not appear, disappear, fragment, or merge between the magnetograms ``propagate'' themselves; they keep their label through all steps.

To track distinct magnetic features, for each flaring active region we identify the features on a reference image composed of ten magnetograms immediately preceding the flare. If some of these (two or more) feature-hills appear as a single feature-hill in most of the frames during the whole analyzed time interval and each of the individual features does not show significant temporal variations, they are merged into one feature on the reference image. An example of an active region with 18 magnetic features found is shown in Figure 1(a). We locate each of these features on each of our images. We analyze 4 h of data covering each flare. We calculate the total flux and the flux-weighted average $x$- and $y$-coordinates of each feature.  

%
\begin{figure}
\centering
\includegraphics[width=1.0\linewidth]{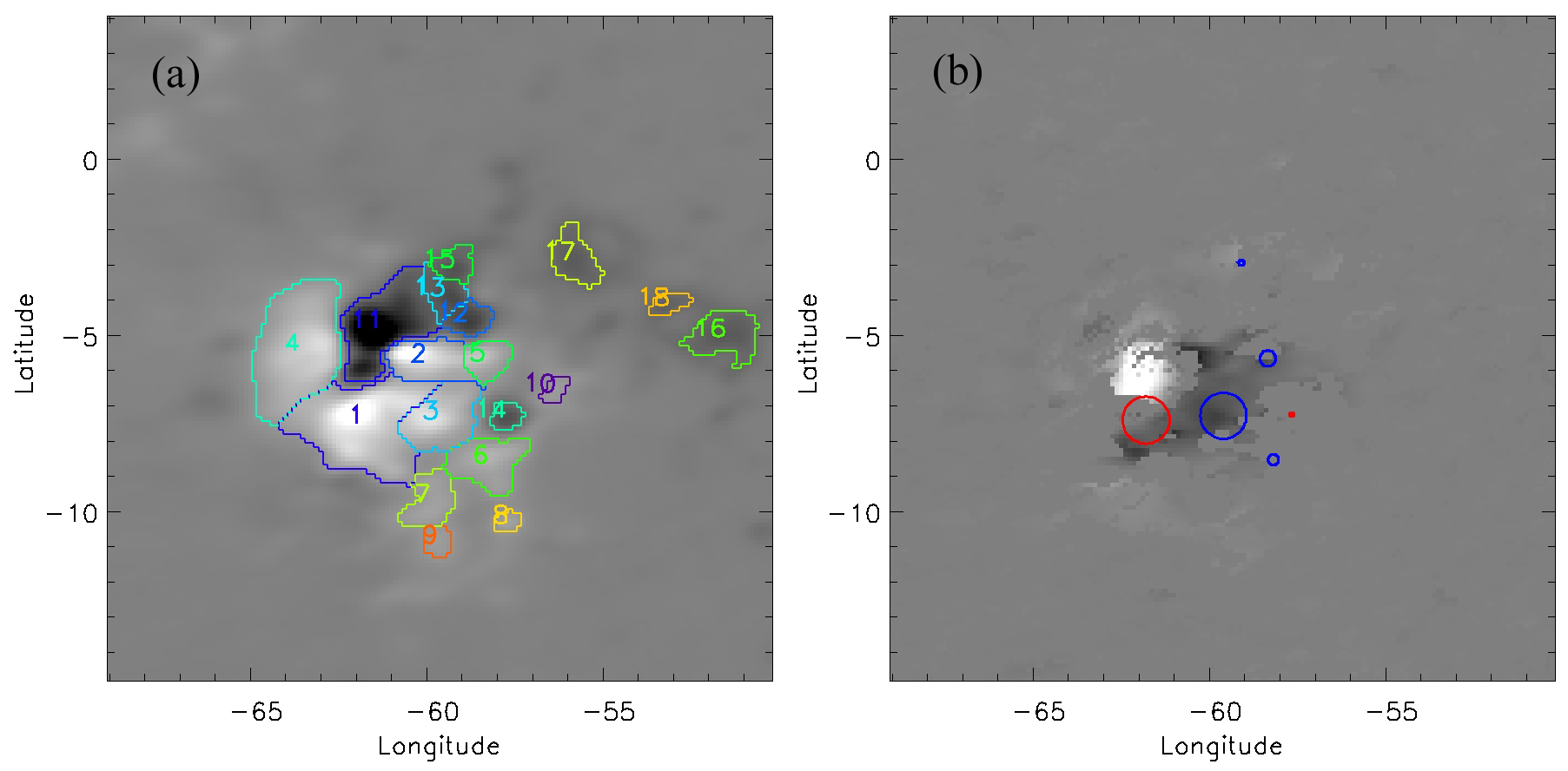}
\caption{(a) Reference image, composed of ten remapped GONG magnetograms, of active region 10930 which produced an X6.5 flare on 6 December 2006 and features identified with the YAFTA downhill algorithm. (b) The $c$-map of stepwise field changes found by \protect \inlinecite{Petrie10} during this flare. The locations of abrupt flux decreases (blue circles) and increases (red circles) of magnetic features found in this work are overplotted.} 
\end{figure}

Most of the previous feature-tracking works have applied a radial field correction to the magnetograms before processing, dividing the longitudinal fields by the cosine of the heliocentric angle ({\it e.g.} \opencite{Welsch11}). For most photospheric fields this is generally a valid assumption \cite{Svalgaard78,Petrie09}. However, for the photospheric fields in a flaring active region, particularly those fields associated with strong, twisted magnetic structures directly involved in the eruption, this assumption may not be applicable. \inlinecite{Petrie10} reported evidence of this in the result that the average magnitude of photospheric field changes tends to increase as a function of distance from the disk center, indicating that the fields changing the most tend to be nearly horizontal. The radial field correction is necessary when features inside of an active region are tracked for a long period of time to ensure a constant threshold during the whole tracking interval ({\it e.g.}, see \opencite{Welsch11}), while we track features only up to 4 h.

\section{Flux Changes}
We analyze the magnitude and sign of flux changes in each magnetic feature during a flare. The magnitude and sign of flux changes are defined by fitting the time evolution of the total flux in each magnetic feature from before the onset to the end of a flare with a straight line, assuming that the flux evolution around the flare is linear and ignoring long-term variations and trends. Figure 2 demonstrates an example of the strongest field change found in our analysis. The spread in the points is mainly caused by seeing changes in the Earth's atmosphere with some contribution from ongoing solar field changes.  

%
\begin{figure}
\centering
\includegraphics[width=0.70\linewidth]{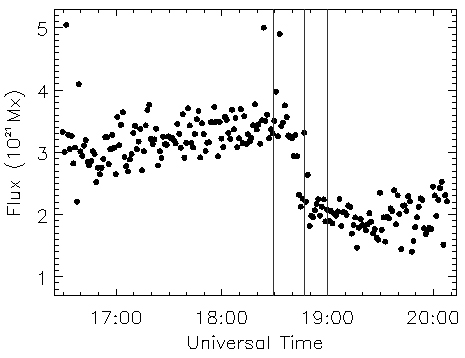}
\caption{Evolution of the magnetic flux in one of the features recorded in AR10930 during an X6.5 flare on 6 December 2006. The three vertical lines indicate the GOES flare start, maximum, and end times.} 
\end{figure}

We selected magnetic features showing stepwise flux changes during flares. Each of the flux changes satisfies the following criterion. The value of the flux change should exceed 1.3 $\sigma$ of the scatter in the measured background flux. The background flux is defined as a median value of the pre-flare flux in the feature. The minimal and maximal values of the background fluxes we found are $2.4\times10^{19}$ and $7.3\times10^{21}$ Mx (maxwell). The 1 $\sigma$ level is between $1.8\times10^{18}$ and $4.5\times10^{20}$ Mx. The maximal and median values of all the stepwise flux changes are 7.3 $\sigma$ and 3.0 $\sigma$, respectively. It has been noted that the duration of the flux changes is on the scale of 10--40 min ({\it e.g.} \opencite{Kosovichev01}; \opencite{Wang10}; \opencite{Petrie10}). We select flux changes which are completed within 30 min or less. 

The statistics for the flux increases and decreases for the 77 flares are presented in Table 1. In total 114 stepwise flux changes which satisfied the above criteria were selected. We find that the field decreases are significantly more numerous than the increases (70 versus 44; see the first two columns under the `All' heading in Table 1). The bias towards the field decreases is noticeable for both disk-center and limb locations with a slightly larger ratio for the limb regions (the first two columns under `Disk-center' and `Limb' headings in Table 1). We designate ``near disk-center'' for the locations $r \le R/2$ and ``near limb'' for the locations $r > R/2$, where $R$ is the solar disk radius in the image plane. Median longitudinal field changes are larger in the regions close to the limb (compare the last two columns under the `Disk-center' and `Limb' headings in Table 1), indicating that their field vector is nearly horizontal in general. Limb changes are more than twice as numerous as disk-center changes. X-class flares show more numerous (74 versus 40) and strong photospheric flux changes in comparison to M-class flares (adding the numbers of flux increases and decreases under the `X-class' and `M-class' headings in Table 1). For X-class flares, the flux decreases are more than twice as numerous as the increases in both disk-center and limb regions, while for M-class flares the ratio of decreases to increases is close to 1:1 for disk-center regions and has an opposite bias (1:2) towards flux increases in regions close to the limb.

\begin{table} 
\caption{The number and the absolute values of abrupt flux changes calculated for individual magnetic features. The values of flux changes are in units of $10^{20}$ Mx.}
\label{T-simple}
\begin{tabular}{@{} l @{} *{5}{c} @{}}     
  \hline                   
Field       & \multicolumn{2}{c} {Number of flux}  & Minimum & Maximum & Median    \\
changes &  increases    &         decreases          & change     & change    & change     \\
  \hline
All                               &   44   &             70              &         0.02   &       21.04    &       0.61        \\
X-class                      &    21   &            53              &         0.02   &       21.04    &      0.81        \\
M-class                      &   23   &            17              &          0.10   &        4.10    &       0.48        \\
Disk-center               &   15   &             22              &         0.10   &        1.84    &       0.35        \\
Limb                           &   29   &            48              &          0.02   &       21.04    &      0.82         \\
Disk-center X-class &     4   &             11             &          0.11   &         1.84    &      0.42         \\
Disk-center M-class &   11   &            11             &          0.10   &         1.60    &      0.27         \\
Limb X-class             &   17   &            42             &          0.02   &       21.04    &      0.82         \\
Limb M-class            &   12   &              6              &          0.15   &         4.10    &      0.84         \\
  \hline
\end{tabular}
\end{table}

These statistics for magnetic features agree with the basic result reported by \inlinecite{Petrie10} that decreasing longitudinal fields tend to outnumber increasing longitudinal fields, except that here magnetic features of near-limb M-class flares have twice as many longitudinal flux increases as decreases. 
   
%
\begin{figure}
\centering
\includegraphics[width=1.0\linewidth]{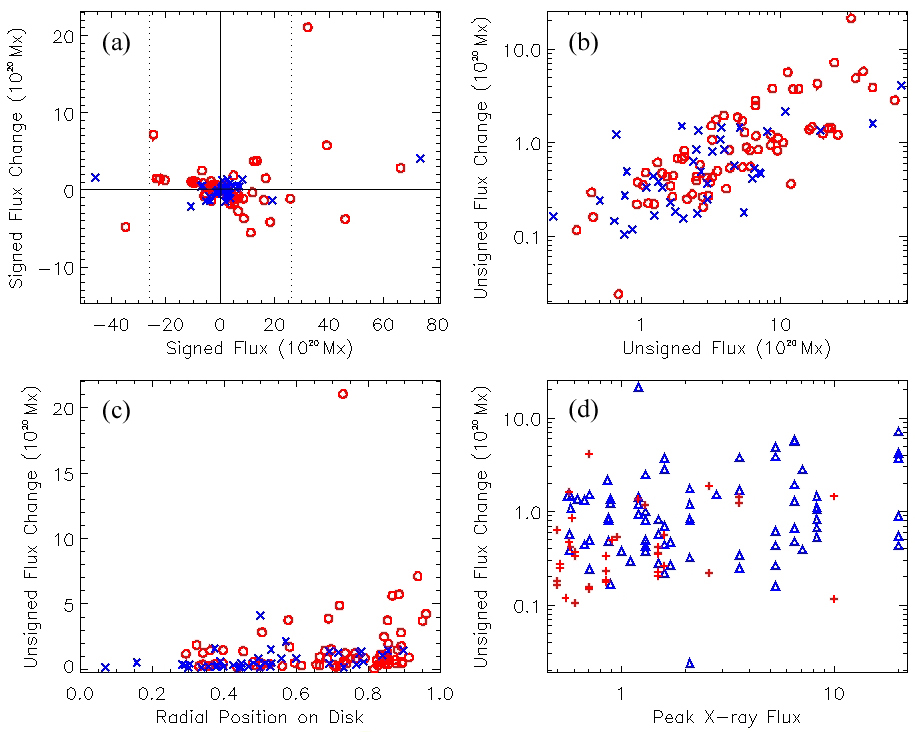}
\caption{Scatter plots of (a) the change in the signed total magnetic flux against the background signed total magnetic flux, and the change in the unsigned total magnetic flux (b) against the background unsigned total magnetic flux and (c) against the radial position of the region on the solar disk during M- (blue crosses) and X-class (red circles) flares. Panel (d) shows a scatter plot of the changes in the unsigned total magnetic flux as a function of the GOES peak X-ray flux for active regions that produced flares near the disk center (red pluses) and near the limb (blue triangles). The flux changes are calculated for individual magnetic features. The dotted lines at $\pm26\times10^{20}$ Mx in panel (a) separate features whose flux change was anti-correlated with their background flux from the big features whose flux change did not conform to this general pattern.} 
\end{figure}

The overall bias towards flux decreases in comparison with flux increases can be seen in the scatter plot of the change in the signed total magnetic flux against the background signed total magnetic flux (Figure 3(a)). The flux changes are calculated for individual magnetic features. In Figure 3, we designate a magnetic flux change ``signed'' if magnetic polarity is taken into account and ``unsigned'' if it is ignored. Each feature consists of pixels of like polarity; thus the flux is total flux in all the features. The strongest flux change was found during an X1.2 flare on 26 October 2003 with its onset at 05:57UT, which is much larger than all other flux changes found in this work. On the other hand, the change in the unsigned total magnetic flux against the background unsigned total magnetic flux (Figure 3(b)) demonstrates that this case of maximum flux change is consistent with the common trend in flux changes, showing that this value may not be an outlier. However, this is a log-log plot. This data point has the same order of magnitude as the rest, but otherwise is an outlier. In any case, large flux changes tend to be associated with X-class flares. Figure 3(c) shows the change in the unsigned total magnetic flux against the radial position of the flare on the solar disk. We can see that the large flux changes associated with X-class flares occur closer to the limb. In total, a larger range of flux changes is found in a flare near the limb than near the disk center. These conclusions are consistent with the statistics of flux changes from \citeauthor{Petrie10} (\citeyear{Petrie10}; see Sections 5 and 7 of their paper). Figure 3(d) shows the change in the unsigned total magnetic flux as a function of the GOES peak X-ray flux. We see a weak overall increasing trend of the flux change with respect to the flare strength.       

We note that the size of magnetic features changes in time, causing an uncertainty in the flux estimates. Some fields crossing the 100 G threshold introduce this error into the flux estimate. However, such errors do not change the basic patterns of increasing and decreasing fluxes.  

\begin{table} 
\caption{Pearson's correlation coefficients for the magnetic flux changes against the background magnetic flux and the GOES peak X-ray flux. The flux changes are calculated for individual magnetic features. The $P$-values are given in parentheses.}
\label{T-simple}
\begin{tabular}{@{} l  *{4}{c} @{}}     
  \hline                   
                                     &             Flux change vs.              &               Flux change vs.             \\
                                     &            background flux              &               peak X-ray flux               \\                          
  \hline
All                                &   -0.42 ($6.6\times10^{-6}$)      &     0.38 ($2.5\times10^{-5}$)       \\
X-class                       &   -0.50 ($1.0\times10^{-5}$)      &     0.35 ($2.4\times10^{-3}$)      \\
M-class                       &                 0.22 (0.19)                   &                0.07 (0.67)                     \\
Disk-center                &   -0.64 ($1.5\times10^{-5}$)      &                0.13 (0.46)                     \\
Limb                            &   -0.40 ($4.1\times10^{-4}$)      &    0.36 ($1.5\times10^{-3}$)       \\
Disk-center X-class  &   -0.74 ($1.0\times10^{-3}$)      &               0.14 (0.64)                      \\
Disk-center M-class  &                0.10 (0.68)                    &               0.02 (0.93)                      \\
Limb X-class              &   -0.49 ($1.3\times10^{-4}$)      &    0.33 ($1.1\times10^{-2}$)       \\
Limb M-class             &                 0.23 (0.37)                   &              -0.01 (0.96)                       \\
  \hline
\end{tabular}
\end{table}

We then compute Pearson's correlation coefficients between the magnetic flux change and the background magnetic flux. We find that the flux change in the features whose absolute value of the background flux is $\le 26\times10^{20}$ Mx is anti-correlated with their background flux, while the larger flux features do not conform to this general pattern (see Figure 3(a)). This bimodal behavior seems to be analogous to the dichotomy between weak ($< 100$ G) and strong ($> 100$ G) field changes found in \inlinecite{Petrie10}. The correlation coefficients computed for magnetic features with absolute values of the background flux $\le 26\times10^{20}$ Mx are shown in Table 2. The $P$-value, the probability that the observed correlation occurs by chance, is also included in this table. We find a modest anti-correlation between the flux change and the background flux for both disk-center and limb locations. The correlation is stronger for near disk-center locations and is dominated by X-class flares. M-class flares show no correlation between the flux change and the background flux; in this case we find about equal numbers of flux decreases and increases (see Table 1).    
 
Pearson's correlation coefficients between the magnetic flux change and the GOES peak X-ray flux show a weak but statistically significant overall correlation dominated by X-class flares at limb locations (see Table 2).  In Table 2, the GOES peak X-ray flux and the position on the disk are approximately equally influential in the correlations.

\subsection{Flux Changes Calculated for Individual Flares}
The magnetic flux changes estimated here represent flux changes at certain locations (in individual magnetic features) within active regions. To obtain a fuller picture of the flare effect on active regions, \inlinecite{Petrie10} summed pixel-wise field changes over the entire active region. We calculate the flux change in a flare by summing abrupt flux changes in the magnetic features during the flare in the active region. Some magnetic features in our analysis did not show stepwise flux changes. The summation of the flux changes over many magnetic features may cause a mixture of some pixels that do not change their flux in an abrupt, stepwise way and some pixels with abrupt, stepwise changes in the flux of opposite signs. This can cause even those magnetic features with many pixels changing stepwise to result in an overall flux profile that does not show a clear stepwise change. We restrict our attention to those magnetic features whose overall flux shows a clear, stepwise, permanent change. We expect these changes to dominate the flux evolution of an active region during a flare.

An example of the $c$-map (a map of stepwise field changes defined by \inlinecite{Petrie10}) is given in Figure 1(b) for the X6.5 flare on 6 December 2006. The locations of magnetic features with abrupt flux changes found in this work are overplotted on the same figure. Feature 3 in Figure 1(a), which demonstrated one of the strongest negative stepwise flux changes in this work (see Figure 2), is fully populated by pixels with strong stepwise changes of the same sign during the flare. Feature 1 included both field decreases and increases. Although most of its pixels did not exhibit clear, stepwise field changes because of spectral line changes during the flare, the summed flux change showed a clear stepwise flux increase. Feature 5 also included pixels with field decreases, and showed a flux decrease twice weaker than that of Feature 3. Feature 2 partially consisted of pixels with the strongest negative field changes, but its flux change did not exceed 1.3 $\sigma$ of the scatter in the background flux, which is the criterion we adopted at the beginning of this section. The other features had smaller stepwise field decreases, and are not specified in Figure 1(a). They also showed stepwise flux changes twice smaller than those of Feature 3. For similar reasons, some of the 77 flares did not include magnetic features whose flux changed in a clear stepwise manner even though each of these flares did include at least one pixel that showed a stepwise flux change \cite{Petrie10}. Also, the imposed threshold for the flux changes has restricted our analysis to strong field regions, which may have left some weaker flux changes in magnetic features beyond our analysis.     

Thus, among 77 flares analyzed, only in 45 (25 X- and 20 M-class) flares we found at least one magnetic feature with strong stepwise flux change. The location of the active region which did not involve stepwise flux changes during some flares does not seem to be correlated with the distance of the active region from the disk center. Some flares in the same active region and at a similar separation from the disk center involved magnetic features with stepwise flux changes and others did not. The detection of the stepwise flux changes did not vary between GONG stations. We note that all cases of the absence of the stepwise flux changes are associated with flare class X1.8 and less, except for one X3.8 flare on 17 January 2005 that occurred near the disk-center location. On the other hand, among 45 flares showing stepwise flux changes, only 12 were above flare class X1.8.       
                             
\begin{table}[t] 
\caption{Net flux changes during flares calculated for individual flares. The values of flux changes are in units of $10^{20}$ Mx.}
\label{T-simple}
\begin{tabular}{@{} l @{} *{5}{c} @{}}     
  \hline                   
Field       & \multicolumn{2}{c} {Number of field} & Minimum & Maximum & Median     \\
changes &     increases       &       decreases       & change     & change    & change      \\
  \hline
All                               &    18   &            27               &        0.002  &      22.17   &       1.17     \\
X-class                      &      8   &             17               &        0.002  &      22.17   &       1.63     \\
M-class                      &   10   &             10               &          0.12   &       2.52    &       0.80     \\
Disk-center               &      5   &             13               &        0.002  &        2.65    &       0.63     \\
Limb                           &   13   &             14               &          0.29   &      22.17   &        1.75     \\
Disk-center X-class &     2    &               5               &        0.002  &        2.65    &        1.32     \\
Disk-center M-class &    3    &               8               &          0.12   &        2.07    &        0.55     \\
Limb X-class             &     6   &             12               &          0.29    &      22.17   &        2.89     \\
Limb M-class            &     7    &              2                &          0.30   &        2.52    &        1.36     \\
  \hline
\end{tabular}
\end{table}

\begin{table}[t] 
\caption{Total flux changes during flares calculated for individual flares. The values of flux changes are in units of $10^{20}$ Mx.}
\label{T-simple}
\begin{tabular}{@{} l @{} *{5}{c} @{}}     
  \hline                   
Field       & \multicolumn{2}{c} {Number of field} & Minimum & Maximum & Median     \\
changes &     increases       &      decreases        & change     & change    & change      \\
  \hline
All                               &   19   &              26              &          0.04  &     17.50    &       1.16      \\
X-class                      &      6   &              19              &         0.29   &     17.50    &       1.75      \\
M-class                      &   13   &                7              &          0.04   &      2.16     &       0.63      \\
Disk-center               &     6    &             12               &         0.07   &      2.65     &       0.84      \\
Limb                           &   13   &              14              &         0.04   &     17.50    &       1.75      \\
Disk-center X-class &     1    &               6               &         0.37   &       2.65    &       1.32      \\
Disk-center M-class &    5    &                6              &         0.07   &       2.16    &        0.55      \\
Limb X-class             &     5   &              13              &         0.29   &     17.50    &        2.73      \\
Limb M-class            &     8   &                 1              &         0.04   &       2.14    &        1.36      \\
  \hline
\end{tabular}
\end{table}

The statistics of net (signed) and total (unsigned) flux changes are shown in Tables 3 and 4, respectively. We find generally more decreases than increases in both net and total flux changes (Tables 3 and 4 under the `All' heading). The bias between the flux decreases and increases is significant in the case of disk-center flares for both net and total flux changes; the ratios of the flux decreases to increases are 2.6:1 and 2:1 for net and total flux changes, respectively (Tables 3 and 4 under `Disk-center' heading). Limb flares show approximately equal numbers of increases and decreases for both net and total flux changes (Tables 3 and 4 under `Limb' heading). 

\inlinecite{Petrie10} calculated the longitudinal field component associated with tilt and azimuth angles of the magnetic vector at different locations on the solar disk. They found that if the field vector becomes more horizontal during a flare, then the longitudinal field can either increase or decrease, both near the disk center or near the limb. However, decreases would likely outnumber increases at all parts of the disk investigated ({\it i.e.}, within $65^\circ$ from the disk center), and this bias is expected to be greater near the disk center than near the limb. We find these patterns in our data.    

For the X-class flares, the flux decreases are at least twice as numerous as the increases at both the disk center and limb locations and for both net and total flux changes (see Tables 3 and 4 under `X-class', `Disk-center X-class', and `Limb X-class' headings). \inlinecite{Petrie10} reported a similar trend for the total flux changes in X-class flares; for the net flux changes they did not find any significant trend. We find more than twice as many flux decreases as increases for the net flux changes in M-class flares at the disk-center locations, which is consistent with the same statistics from \inlinecite{Petrie10}, while the trend is opposite for the limb flares (Table 3 under `M-class', `Disk-center M-class', and `Limb M-class' headings). For the total flux changes in M-class flares, the increases are nearly twice as numerous as the decreases in general (Table 4 under `M-class' heading). A significant bias towards total flux increases in near-limb M-class flares is observed, while the same category from \inlinecite{Petrie10} does not show this trend. Similar to the feature-wise statistics, the median flux changes are larger in the regions close to the limb for both total and net flux changes and larger in the case of X-class flares in comparison with M-class flares.

%
\begin{figure}
\centering
\includegraphics[width=1.0\linewidth]{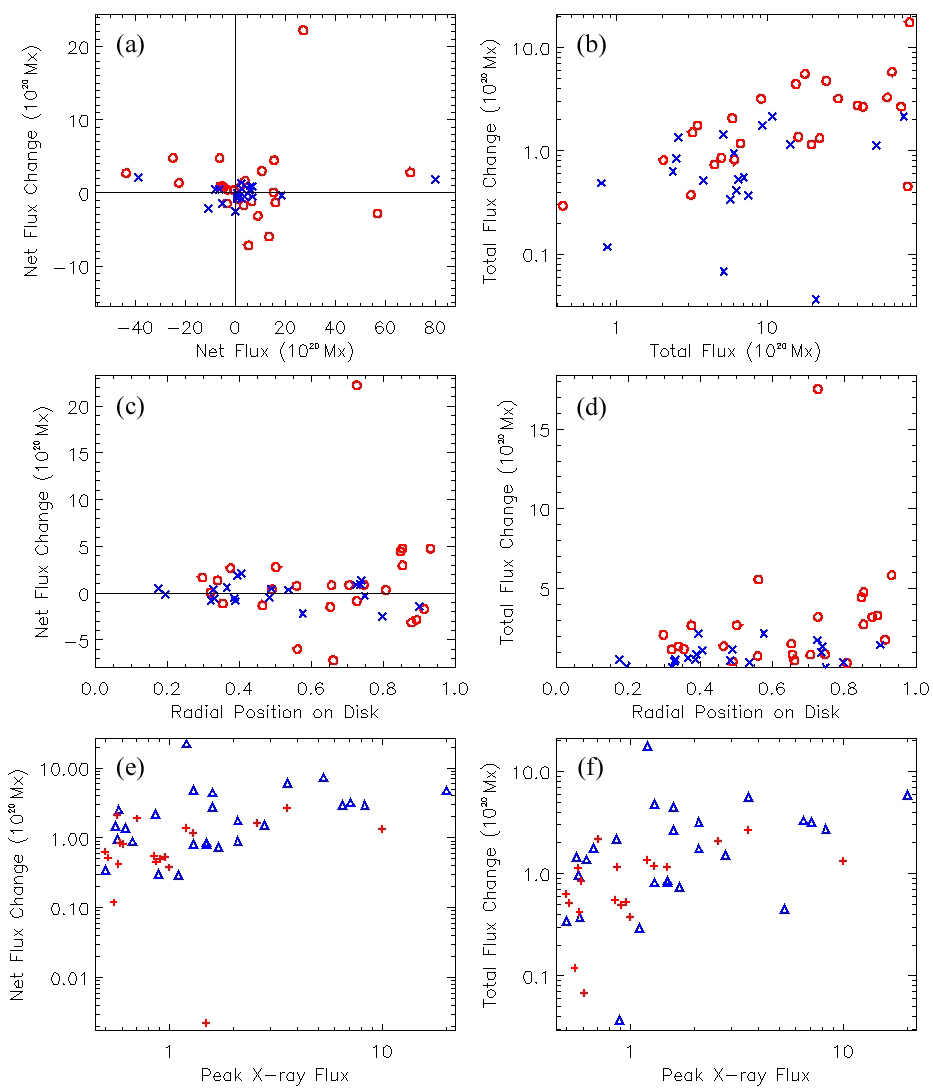}
\caption{Scatter plots of the changes in the net magnetic flux against (a) the background net flux, (c) the radial position of the region on the solar disk for M- (blue crosses) and X-class (red circles) flares, and (e) the GOES peak X-ray flux for active regions near the disk center (red pluses) and near the limb (blue triangles), respectively. Panels (b), (d), and (f) are similar plots for the changes in the total magnetic flux. The flux changes are calculated for individual flares.} 
\end{figure}
%

\begin{table} 
\caption{Pearson's correlation coefficients for the magnetic flux change against the background magnetic flux calculated for individual flares. The $P$-values are given in parentheses.}
\label{T-simple}
\begin{tabular}{@{} l  *{4}{c} @{}}     
  \hline                   
                                  &                        \multicolumn{2}{c} {Background flux vs.}                \\
                                  &                      net flux                      &                 total flux                    \\
                                  &                      change                     &                  change                     \\                                  
  \hline
All                               &                  -0.28 (0.09)                &   0.55 ($7.0\times10^{-5}$)    \\
X-class                      &                  -0.39 (0.08)                &    0.55 ($4.0\times10^{-3}$)    \\
M-class                      &                   0.32 (0.20)                &   0.47 ($3.4\times10^{-2}$)     \\
Disk-center               &    -0.56 ($2.4\times10^{-2}$)    &   0.68 ($1.0\times10^{-3}$)     \\
Limb                           &                 -0.26 (0.23)                 &   0.56 ($2.3\times10^{-3}$)     \\
Disk-center X-class &                 -0.63 (0.19)                 &               0.68 (0.10)                   \\
Disk-center M-class &                -0.42 (0.24)                 &   0.87 ($5.5\times10^{-5}$)      \\
Limb X-class             &                -0.41 (0.13)                 &   0.52 ($2.5\times10^{-2}$)      \\
Limb M-class            &                  0.55 (0.17)                 &              -0.33 (0.44)                   \\
  \hline
\end{tabular}
\end{table}

\begin{table} 
\caption{Pearson's correlation coefficients for the magnetic flux change against the GOES peak X-ray flux calculated for individual flares. The $P$-values are given in parentheses.}
\label{T-simple}
\begin{tabular}{@{} l  *{4}{c} @{}}     
  \hline                   
                                  &                    \multicolumn{2}{c} {Peak X-ray flux vs.}                 \\
                                  &                 net flux                       &              total flux                      \\
                                  &                 change                      &               change                       \\                                  
  \hline
All                               &   0.49 ($6.5\times10^{-4}$)   &  0.56 ($8.5\times10^{-5}$)    \\
X-class                      &                 0.39 (0.07)               &  0.47 ($1.8\times10^{-2}$)     \\
M-class                      &               -0.11 (0.65)               &            0.21 (0.40)                     \\
Disk-center               &                0.34 (0.17)                &            0.36 (0.14)                     \\
Limb                           &   0.48 ($1.4\times10^{-2}$)   &  0.56 ($3.9\times10^{-3}$)     \\
Disk-center X-class &                0.20 (0.72)                &            0.20 (0.68)                     \\
Disk-center M-class &              -0.15 (0.65)                &            0.09 (0.80)                     \\
Limb X-class             &                0.40 (0.11)                &  0.50 ($4.1\times10^{-2}$)     \\
Limb M-class            &               0.003 (0.99)               &  0.81 ($2.2\times10^{-2}$)     \\
  \hline
\end{tabular}
\end{table}

Figure 4 shows scatter plots of the changes in the net and total magnetic flux against the background net and total magnetic flux, against the radial position of the region on the solar disk, and against the GOES peak X-ray flux. The correlations between the magnetic flux changes and the background magnetic flux and the GOES peak X-ray flux are summarized in Tables 5 and 6. The only statistically significant modest anti-correlation between the net flux change and the net background flux is found for the disk-center flares (see Table 5 under `Disk-center' heading), which agrees with the results in Table 3. Some weak anti-correlation is also noted for the X-class flares (Table 5 under `X-class' heading, first column), which is consistent with Figure 4(a) and the statistics in Table 3. The distribution of points in Figure 4 is not much different from the one shown in Figure 3. Figure 4(b) shows more points that do not conform to the general trend for the total flux changes in M-class flares, while the flux changes in X-class flares maintain this pattern quite well, except for one point corresponding to one of the limb flares. Figures 4(c) and 4(d) clearly show that stronger flux changes occur preferentially closer to the limb and that a larger range of the flux changes corresponds to the limb locations.   

Figures 4(e) and 4(f) show an overall increasing trend of the net and total flux changes as a function of the peak X-ray flux. We do find moderately significant correlation with the peak X-ray flux for both net and total flux changes, which are dominated by the X-class flares near the limb locations. This tendency is clearer for the total flux changes; the net flux changes show statistically significant correlation only for limb flares (compare Table 6 under `All' vs. `Limb' and `X-class' vs. `Limb X-class').

\section{Flux Cancellation}
We also identify episodes of flux cancellation, when closely spaced, opposite-polarity features simultaneously lose flux \cite{Livi85,Martin85}. We estimate the flux cancellation as described in \inlinecite{Welsch11}. The magnetic flux of features derived from high-cadence data suffers short-term fluctuations (noise) with large amplitudes in this study, as the GONG data are also influenced by the Earth's atmosphere. The fluctuations significantly affect the measurements of flux cancellation, leading to exaggerated values of the flux losses in some cases by summing up noisy data instead of real flux evolution. We smooth the total flux in each feature with a 30-min running mean. We then mostly see a long-term evolutionary trend of the flux that we use in further investigations of flux cancellation. First, we define instances when the boundaries of closely spaced, opposite-polarity features come into contact, using the technique described in \inlinecite{Welsch11}. For our analysis, the closest distance between the features is limited by the pixel size of the GONG data to about 1.8 Mm on the Sun at disk center. Further, for a pair of neighboring features to be partially canceled in the next magnetogram, the centers of magnetic features are required to approach and the flux of both features to decrease. This approach is subject to many sources of uncertainties, related to the applied threshold (some flux may be missed by the threshold), fluctuations in magnetic flux due to measurement noise and true evolution, as well as possible interaction among several neighboring like- or opposite-polarity features. To account for these uncertainties, flux cancellation is computed in three ways as in \inlinecite{Welsch11}: 
\begin{itemize} 
\item[-] estimated minimum value of flux loss from the pair of canceling features, summed over all cancellation events;
\item[-] estimated average value of flux loss from the pair of features exhibiting flux cancellation over at least two consecutive magnetograms (multi-step cancellation), summed over all multi-step cancellation events;
\item[-] estimated minimum value of flux loss from the pair of features exhibiting flux cancellation over at least two consecutive magnetograms, summed over all multi-step cancellation events.
\end{itemize}

\begin{table}[ph!] 
\caption{The amount of canceled magnetic flux (in units of $10^{19}$ Mx) for a 4-h interval around each M-class flare, computed in three ways (see text).}
\label{T-simple}
\begin{tabular}{@{} l *{7}{c} @{}}     
  \hline                   
Date                &       GOES          &  Radial position   &  Number of   &    Number of          &          All events           &                Multi-step         &            Multi-step  \\
                         &        class           &   (in $R_\odot$)   &      events      & multi-step events &  $\sum {\rm min}(\delta\Phi)$  &  $\sum {\rm avg}(\delta\Phi)$ &   $\sum {\rm min}(\delta\Phi)$  \\
  \hline
22 Jun 2001  &         M6.2           &            0.74           &           -              &            -               &                   -                      &                  -                       &                   -                        \\      
23 Jun 2001  &         M5.6           &            0.43           &          29            &             2             &                   5.5                  &                 0.2                    &                  0.2                    \\ 
05 Sep 2001  &         M6.0           &            0.49           &             9           &             -             &                   0.7                   &                  -                       &                    -                      \\ 
09 Sep 2001  &         M9.5           &            0.70           &           14           &             -             &                  1.1                   &                   -                       &                    -                      \\ 
16 Sep 2001  &         M5.6           &            0.90          &             2           &             -              &                 0.4                   &                   -                        &                   -                       \\ 
22 Oct 2001   &         M6.7           &            0.52          &            35          &            8             &                  2.2                   &                1.3                      &                 0.6                    \\ 
23 Oct 2001   &         M6.5            &            0.43          &            39          &           13             &                 5.0                   &                5.7                     &                 2.5                     \\ 
07 Nov 2001  &         M5.7           &            0.74           &           96           &           21             &               18.2                   &               7.1                     &                 4.6                    \\ 
08 Nov 2001  &         M9.1           &            0.48           &             5           &             -              &                 0.4                    &                -                         &                   -                      \\ 
28 Nov 2001  &         M6.9           &            0.28           &           53           &           13             &                 7.7                    &                3.1                     &                 2.0                   \\ 
29 Nov 2001  &         M5.5            &            0.19          &         133           &           29             &               11.2                   &                4.2                    &                  2.5                   \\ 
26 Dec 2001  &         M7.1           &            0.83           &           26           &             4              &               12.9                   &                4.8                    &                  3.6                   \\ 
09 Jan 2002   &         M9.5            &            0.29          &            60           &             9              &               10.8                   &                3.7                    &                  2.4                   \\ 
14 Mar 2002  &         M5.7            &            0.41          &            28           &             9              &                 3.6                   &                2.7                    &                  1.5                    \\ 
11 Jul 2002   &          M5.8            &            0.80          &              -             &             -               &                  -                       &               -                        &                   -                       \\ 
17 Jul 2002   &          M8.5             &            0.33          &            26           &            7              &                 1.1                    &                1.7                   &                  0.5                    \\ 
26 Jul 2002   &          M8.7             &            0.49          &          173           &          38             &                21.7                   &                9.5                    &                 6.2                    \\          
16 Aug 2002  &        M5.2             &            0.39          &            62           &          11              &               11.7                   &                4.4                   &                  3.0                    \\
05 Oct 2002   &         M5.9             &            0.52           &           10           &             2              &                 0.3                   &                0.2                    &                 0.1                   \\
18 Nov 2002   &        M7.4             &            0.80          &             -             &            -                &                  -                      &                  -                      &                   -                       \\
20 Dec 2002   &        M6.8             &            0.65          &          183           &          52              &               12.2                   &               7.1                   &                 3.9                    \\
26 Oct 2003   &          M7.6            &            0.63          &             95          &           12              &               44.8                  &              13.3                   &                 6.1                    \\
27 Oct 2003   &          M5.0            &            0.54          &          110           &           28             &               11.0                  &               10.8                   &                 3.3                    \\
20 Nov 2003  &         M5.8            &            0.33          &            45           &            9              &                 3.5                   &                1.5                   &                  0.5                    \\  
20 Nov 2003  &         M9.6            &            0.18          &            88           &           28             &                 7.9                   &                6.2                  &                  2.6                    \\
17 Jan 2004  &          M5.0            &            0.37          &             -             &             -               &                -                       &                 -                       &                   -                       \\
20 Jan 2004  &          M6.1            &            0.32          &              7           &             -              &                0.1                   &                  -                       &                   -                      \\
13 Jul 2004  &           M6.2            &            0.83          &            10           &             -              &                0.8                   &                  -                       &                   -                     \\
13 Jul 2004  &           M6.7            &            0.72          &            14           &             2              &                2.5                   &                0.5                   &                0.5                    \\
20 Jul 2004  &           M8.6            &            0.58          &              -            &             -               &                 -                      &                 -                       &                  -                        \\
22 Jul 2004  &           M9.1            &            0.35          &            31           &             2              &                1.3                  &                0.04                   &               0.01                   \\
25 Jul 2004  &           M7.1            &            0.39          &            21           &             6              &                1.5                 &                  1.1                   &                0.4                     \\
14 Aug 2004 &          M7.4            &            0.56          &            43           &           10              &                8.5                 &                  2.5                   &                1.3                     \\  
30 Oct 2004  &           M5.9            &            0.39          &            37           &          10              &                3.1                 &                  4.9                   &                0.6                     \\ 
15 Jan 2005  &          M8.4            &            0.33          &            62           &          12              &              16.9                  &                 5.7                   &                2.0                    \\ 
15 Jan 2005  &          M8.6            &            0.31          &            63           &          12              &              20.8                  &               10.7                   &                5.3                    \\ 
06 Dec 2006  &         M6.0             &            0.90          &             -             &            -               &                -                      &                   -                       &                 -                        \\ 
04 Jun 2007   &          M8.9            &            0.75           &           34           &            7              &                2.4                  &                  1.5                   &                0.5                     \\ 
  \hline
\end{tabular}
\end{table}

\begin{table}[ph!]
\caption{The amount of canceled flux (in units of $10^{19}$ Mx) for a 4-h interval around each X-class flare, computed in three ways (see text).}
\label{T-simple}
\begin{tabular}{@{} l *{7}{c} @{}}     
  \hline                   
Date                &       GOES          &  Radial position   &  Number of   &    Number of          &           All events        &           Multi-step               &            Multi-step   \\
                         &        class           &   (in $R_\odot$)   &      events      & multi-step events  &  $\sum {\rm min}(\delta\Phi)$  &  $\sum {\rm avg}(\delta\Phi)$ &   $\sum {\rm min}(\delta\Phi)$  \\
  \hline
02 Apr 2001  &          X20.0          &            0.93          &        54           &          17             &                  63.5                  &              23.9                     &                17.0                      \\
23 Jun 2001  &          X1.2            &            0.41          &        42           &          12             &                    5.9                  &                2.3                     &                  1.3                      \\                        
25 Aug 2001 &          X5.3            &            0.66          &         58           &          10             &                    2.7                  &                0.9                     &                  0.4                       \\
19 Oct 2001   &         X1.6             &            0.50          &        27           &            7             &                    1.8                  &                 0.8                     &                  0.5                      \\
22 Oct 2001   &         X1.2             &            0.47          &        26           &            9              &                   5.8                  &                 4.8                     &                  2.4                      \\         
11 Dec 2001  &         X2.8             &            0.65          &          8           &            2              &                   0.8                   &                0.9                     &                  0.3                      \\          
20 May 2002  &         X2.1             &            0.91          &          2           &            -              &                   0.2                   &                  -                        &                   -                        \\        
21 Aug 2002  &         X1.0             &            0.72          &        19           &            6              &                   5.9                   &                3.8                     &                 3.0                      \\         
17 Mar 2003  &         X1.5             &            0.59          &         70           &          26             &                 28.6                   &              17.6                    &               12.5                      \\      
18 Mar 2003  &         X1.5             &            0.71          &       113          &           31             &                 24.2                   &              10.7                    &                 6.6                      \\      
27 May 2003  &        X1.3             &            0.36          &         72           &           11             &                 11.2                   &                3.8                    &                 2.4                      \\      
28 May 2003  &        X3.6              &           0.38          &        84           &           17             &                 12.1                   &                 6.1                    &                 3.0                      \\      
10 Jun 2003   &         X1.3             &            0.66          &         96           &           14             &                 11.7                   &                3.7                    &                1.7                      \\     
11 Jun 2003   &         X1.6             &            0.85          &         77           &           11             &                 14.5                   &                3.2                    &                1.6                      \\      
19 Oct 2003   &          X1.1             &            0.81          &         14           &             -              &                   4.5                   &                 -                        &                  -                        \\       
26 Oct 2003  &          X1.2a           &            0.73          &         84           &           17             &                 25.1                   &             16.4                    &                6.0                      \\     
26 Oct 2003  &          X1.2b           &            0.62          &       127           &           16             &                 33.3                  &              15.2                    &                4.0                      \\      
29 Oct 2003  &          X10.0           &            0.34          &       153           &           37             &                 40.4                  &              25.6                    &              14.4                      \\
02 Nov 2003 &          X8.3             &            0.85          &          -             &             -               &                   -                       &                 -                       &                 -                           \\
26 Feb 2004  &         X1.1             &            0.43          &         19           &              2             &                   2.7                  &                0.8                    &                 0.4                      \\  
15 Jul 2004  &           X1.6             &            0.72          &         44           &             6             &                 12.4                  &                2.8                    &                 1.5                      \\
15 Jul 2004  &           X1.8             &            0.79          &           2           &              -             &                   0.2                   &                 -                       &                   -                         \\
16 Jul 2004  &           X1.1             &            0.56          &         59           &           10             &                 14.0                  &                2.4                    &                 1.4                      \\
16 Jul 2004  &           X1.3             &            0.63          &         51           &           15             &                 10.6                  &                6.4                    &                 3.2                      \\
16 Jul 2004  &           X3.6             &            0.56          &         71           &           16             &                 18.7                   &               6.0                    &                 3.7                       \\
13 Aug 2004  &         X1.0             &            0.49          &         63           &           16             &                   4.9                   &               1.9                    &                  1.4                      \\
30 Oct 2004  &          X1.2             &            0.40          &         29           &              4             &                   2.2                  &                0.4                    &                   0.1                      \\
01 Jan 2005  &         X1.7              &            0.56          &          -             &             -               &                     -                     &                 -                       &                  -                          \\
15 Jan 2005  &         X1.2              &            0.32          &       194          &            35             &                 16.0                   &               7.5                    &                 2.7                     \\
15 Jan 2005  &         X2.6              &            0.30          &        297          &           95             &                 28.3                   &             19.2                    &                 7.8                      \\
17 Jan 2005  &         X3.8              &            0.46          &        203          &           54             &                 73.2                   &             31.2                    &               17.4                      \\
20 Jan 2005  &         X7.1              &            0.88          &            7          &              -              &                   7.6                   &                 -                       &                  -                         \\
30 Jul 2005   &         X1.3              &            0.86          &            -            &             -              &                    -                      &                  -                       &                  -                          \\
10 Sep 2005  &        X1.1              &            0.75          &          39           &              3            &                   9.8                   &                3.2                    &                0.6                      \\
10 Sep 2005  &        X2.1              &            0.73          &          43           &              8             &                12.5                  &                 4.7                    &                1.9                      \\
13 Sep 2005  &        X1.5              &            0.32          &        120           &           15             &                  9.5                   &                 2.2                   &                 1.1                      \\
15 Sep 2005  &        X1.1              &            0.40          &          90           &           17             &                16.3                   &                5.6                    &                3.7                      \\
06 Dec 2006  &        X6.5              &            0.89          &          60           &           10             &                44.1                   &              14.4                    &                8.0                      \\
14 Dec 2006  &        X1.5              &            0.75          &          48           &             8             &                  7.5                   &                 4.5                   &                 1.8                      \\
  \hline
\end{tabular}
\end{table}

The total number of cancellation events and the estimates of the flux losses during the whole 4-h interval around each flare of M- and X-classes are listed in Tables 7 and 8, respectively. Ordering of all the estimates obtained with the three methods for each flare is consistent with that of \inlinecite{Welsch11}. The third technique is the most restrictive, and thus should be the most reliable, showing a real physical pattern. 

%
\begin{figure}[h!]
\centering
\includegraphics[width=0.93\linewidth]{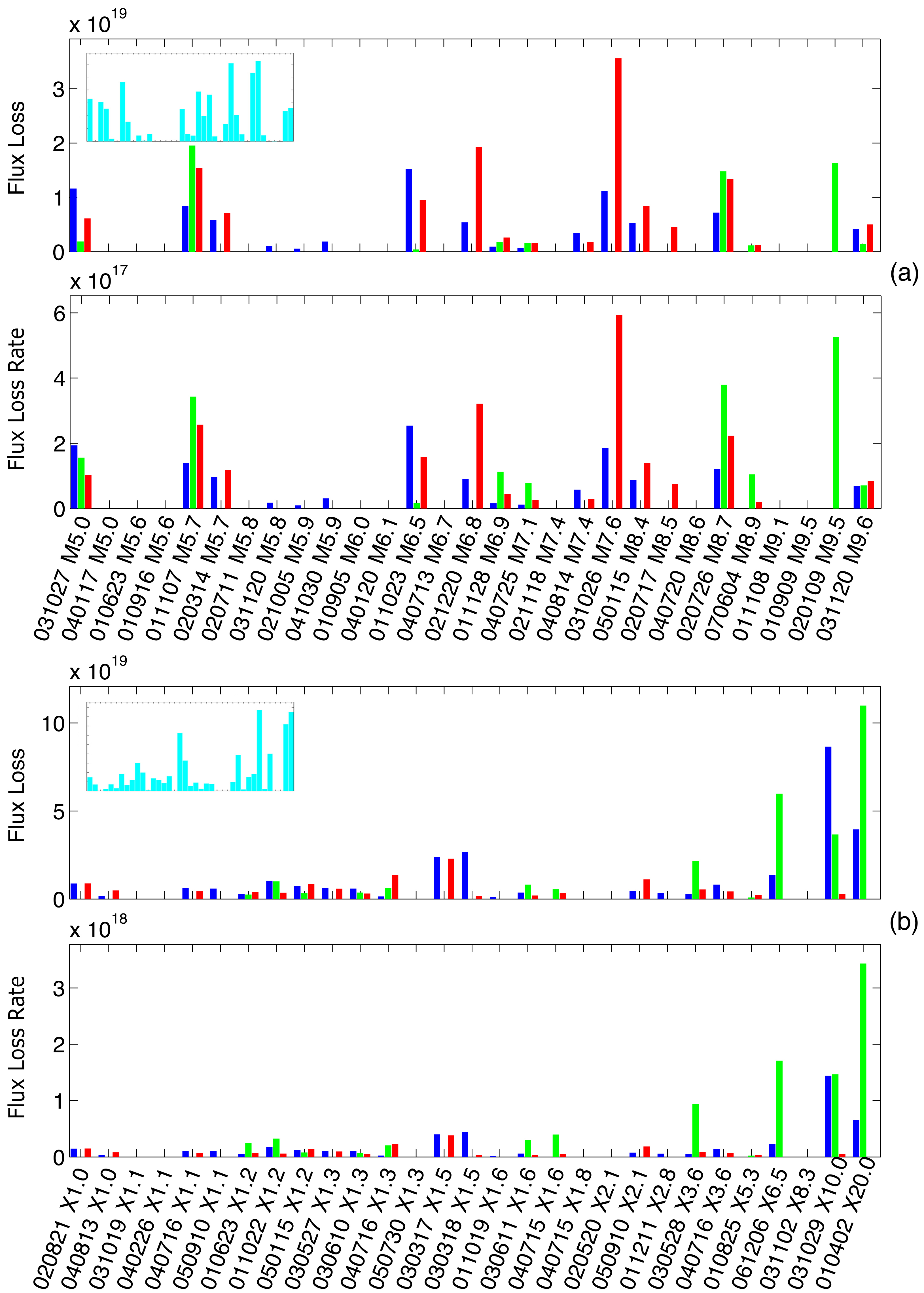}
\caption{Estimated amount of the canceled flux [in Mx] in the active regions over 60 min before (blue), 60 min after (red), and during (green) a flare in multi-step events in which flux cancellation was observed over at least two consecutive magnetograms. The duration of flares varies from 8 to 57 min. Panels (a ) and (b) show the results for M- and X-class flares, respectively. In each panel, the upper graph shows the sum of minimum flux loss for each pair of magnetic features in each event; the lower graph shows the average flux loss rate [in Mx min$^{-1}$] (the amount of lost flux divided by the observed time interval). Flux losses are plotted in ascending order of flare class. The inset plots show the total flux loss over the entire time interval around each flare (one bar per flare).} 
\end{figure}

We also compute the amount of canceled magnetic flux separately for three time intervals: before the flare onset, during the flare, and after the end of the flare. The flare start, peak, and end times are taken from GOES data. By this division of our data sets by three time intervals, we end up with different durations of these intervals for each flare. For a fair comparison, we restrict the time intervals before and after the flare to 60 min each. If a flare data set is shorter in at least one of the two time intervals, we exclude the flare from comparison, thus obtaining a final inclusion of 29 X- and 29 M-class flares. The duration of the flare is different for all flares and varies from 8 to 57 min. Note that the `during-flare' interval is always shorter than 60 min. Flux losses for 29 X- and 29 M-class flares obtained by the third, most restrictive, technique for estimating flux cancellation are shown in Figure 5. We compute mean and median values of flux losses in flux cancellation events before, during, and after the flare for both the amount of flux loss and the average flux loss rate. The results are shown in Table 9. 
     
The amount of flux canceled was greater over the time during the flare than before and after the flare for X-class flares. The ratios of the flux losses before, during, and after the flare to the amount of flux canceled before the flare are 1:3:0.7 (median values) and 1:4:0.5 (mean values). Some of the strongest X-class flares showed significantly greater flux loss during the flare time. The flux loss before the flare was greater than that after the flare by a smaller factor. For the M-class flares no significant trend was found.

\begin{table}[b] 
\caption{Statistics of flux losses and average flux loss rates in flux cancellation events before, during, and after the flare.}
\label{T-simple}
\begin{tabular}{@{} l r @{} *{3}{c} @{}}     
  \hline                   
\multicolumn{2}{c} {}     &  Before flare  &  During flare  & After flare     \\
  \hline
Median flux loss            & Flux loss [$10^{18}$ Mx]                                                         \\  
                                         & M-class flares &          5.2          &         1.8           &       6.6    \\
                                         & X-class flares  &          6.1          &          7.1          &       4.4    \\
 
                                         & Flux loss rate [$10^{18}$ Mx min$^{-1}$]   \\
                                         & M-class flares &    0.09    &    0.1    &    0.1    \\
                                         & X-class flares &     0.1      &    0.3    &    0.07  \\
\hline
Mean flux loss               & Flux loss [$10^{18}$ Mx]   \\
                                         & M-class flares &    5.5    &    6.5    &    9.4      \\
                                         & X-class flares  &  12.9    &  22.3   &    6.3      \\

                                         & Flux loss rate [$10^{18}$ Mx min$^{-1}$]   \\
                                         & M-class flares &    0.09    &    0.2    &    0.16   \\
                                         & X-class flares  &    0.2      &    0.8    &    0.1     \\
  \hline  
\end{tabular}
\end{table}

The flux losses in Figure 5 are plotted in ascending order of flare class. We note an overall increasing trend of the flux losses as a function of flare power, which is more obvious for X-class flares than for M-class flares. This is clear from the inset plots of Figure 5, where each bar represents flux loss over the entire 4-h interval around each flare (from the last columns of Tables 7 and 8) and is plotted in the ascending order of flare class. Can the correlation between the flux loss and the total flux in the canceling features or in the active region account for this trend?

\subsection{Correlation between the Flux Loss and the X-Ray Flux}
Figure 6 shows, as a function of the GOES peak X-ray flux, the amount of magnetic flux loss, total magnetic flux in the active region, the number of canceling magnetic features, and the total magnetic flux in the canceling features, respectively. The Pearson correlation coefficients for all the quantities are summarized in Table 10. The increasing trend as a function of the peak X-ray flux is dominated by X-class flares, as can be seen in Figure 6 for the flux loss and the total flux in the active region but is less obvious for the number of canceling features and the total flux in canceling features. We find that all four quantities show weak-to-moderate correlation with peak X-ray flux at disk-center locations, and the correlation is dominated by X-class flares. The exception is the number of canceling features, which shows only weak correlation with the peak X-ray flux for disk-center flares and no statistically significant correlation for the disk-center X-class flares (Table 10 under `Disk center' and `Disk-center X-class' headings). The correlation is highest for the flux loss, which also shows moderate correlation for the limb flares dominated by the X-class flares (the `Flux loss' column of Table 10). The M-class flares revealed no correlation with the peak X-ray flux in any of the categories. 
   
%
\begin{figure}[h!]
\centering
\includegraphics[width=1.0\linewidth]{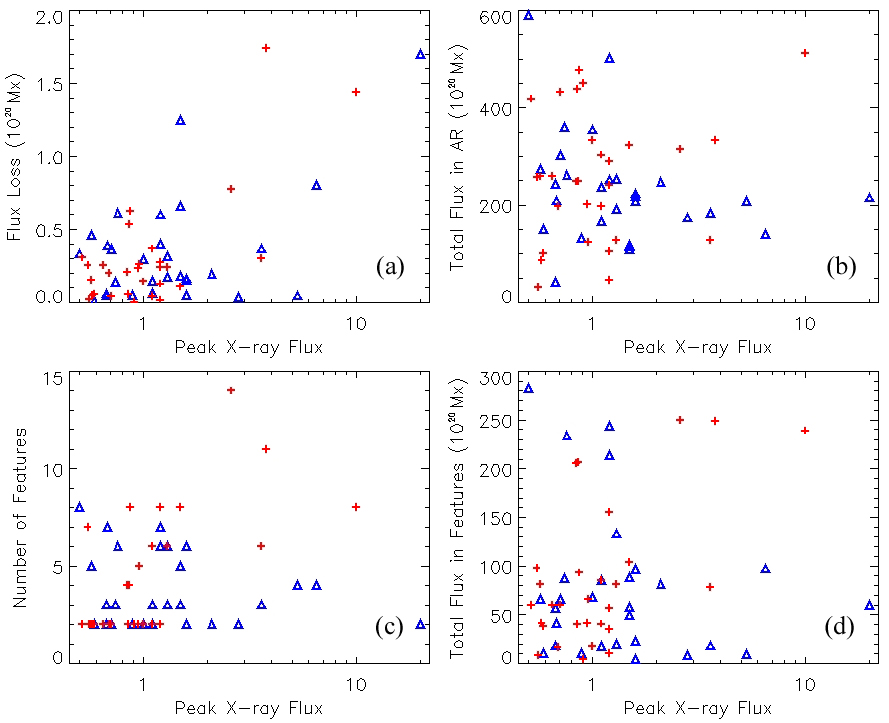}
\caption{(a) The amount of magnetic flux loss, (b) the total magnetic flux in the active region, (c) the number of canceling features, and (d) total flux in the canceling features against the GOES peak X-ray flux for active regions near the disk center (red pluses) and near the limb (blue triangles).} 
\end{figure}

Previous studies found that the total magnetic flux in the active region and the magnetic flux at the vicinity of the polarity inversion line (PIL) correlate with flare activity \cite{Leka07,Schrijver07,Barnes08,Welsch09}. The latter was found to best correlate with the peak X-ray flux \cite{Schrijver07}. We find about the same level of low-to-moderate correlation, dominated by X-class flares at disk-center locations, between the peak X-ray flux and both the magnetic flux in the canceling features and the total flux in the active region. Here, only some of the PIL features contribute to the multi-step flux cancellation events; thus the flux in the canceling features and the flux at the PIL are not exactly the same quantities. The relation between the flux in PIL features and the flare strength will be discussed in the next section.       

We have explored how the amount of magnetic flux loss, the total magnetic flux in the active region, the number of canceling features, and the total flux in the canceling features correlate with each other. The results are shown in Table 11. We find that both the number of canceling features and the total flux in the canceling features show moderate positive correlation with the flux loss in all categories. The exceptions are limb flares, particularly limb X-class flares, which did not show any correlation with the flux loss. However, some moderate correlation is found for the limb M-class flares (see the last two columns in Table 11 under `Limb M-class' heading). The correlation for the flux in the canceling features is dominated by the disk-center X-class flares. In contrast, the total flux in the active region shows weak-to-moderate correlation only with the flux loss for the disk-center flares and with contribution only of the X-class flares. We also note that the flux loss is approximately equally correlated with the number of canceling features and the total flux in these features (see the last two columns in Table 11), indicating that the field strength and complexity are approximately equally important. 

\begin{table}
\caption{Pearson's correlation coefficients for the amount of magnetic flux loss, the total magnetic flux in the active region, the number of canceling features, and the total flux in the canceling features against the GOES peak X-ray flux for active regions near the disk center and near the limb. The $P$-values are given in parentheses.}
\label{T-simple}
\begin{tabular}{@{} l *{4}{c} @{}}     
  \hline                   
                                   &                  Flux loss                    &                Total flux in            &                Number of             &             Total flux in               \\    
                                   &                                                      &              active region           &        canceling features       &      canceling features          \\                                     
  \hline
All                               &   0.66  ($5.8\times10^{-9}$)   &    0.07 (0.57)                          &  0.08  (0.56)                          &  0.09  (0.49)                           \\
X-class                      &   0.66  ($1.8\times10^{-5}$)    &   0.16 (0.32)                          &  0.01 (0.95)                           &  0.10 (0.60)                             \\
M-class                      &   0.12  (0.58)                             &   0.18 (0.29)                           & -0.02 (0.93)                          & -0.02 (0.93)                             \\
Disk-center               &   0.74  ($1.3\times10^{-6}$)   &   0.35 ($4.8\times10^{-2}$) &  0.47 ($8.7\times10^{-3}$) &  0.56 ($1.2\times10^{-3}$)   \\
Limb                           &   0.68  ($2.8\times10^{-5}$)   & -0.01 (0.94)                            & -0.19 (0.33)                          & -0.11 (0.56)                             \\
Disk-center X-class &   0.75  ($2.2\times10^{-3}$)   &   0.61 ($2.5\times10^{-2}$) &   0.39 (0.20)                          &  0.64 ($1.6\times10^{-2}$)   \\
Disk-center M-class &  0.28  (0.30)                              &  0.27 (0.25)                           &   0.24 (0.37)                          &  0.22 (0.43)                             \\
Limb X-class             &  0.70  ($5.5\times10^{-4}$)    &  0.05 (0.81)                           &  -0.22 (0.38)                          & -0.12 (0.63)                             \\
Limb M-class            & -0.14  (0.71)                              &  0.02 (0.92)                           &  -0.44 (0.21)                          & -0.32 (0.38)                             \\
  \hline
\end{tabular}
\end{table}

\begin{table}[b]
\caption{Pearson's correlation coefficients for the amount of magnetic flux loss against the total magnetic flux in the active region, the number of canceling features, and the total flux in the canceling features, respectively. The divisions are made for M- and X-class flares and their locations near the disk center and near the limb. The $P$-values are given in parentheses.}
\label{T-simple}
\begin{tabular}{@{} l *{4}{c} @{}}     
  \hline                   
Flux loss                   &               Total flux in               &                 Number of                &        Total flux in                       \\    
                                   &             active region              &          canceling features         &   canceling features                \\                                     
  \hline
All                                &  0.21 (0.11)                           &    0.47 ($1.4\times10^{-4}$)   &    0.50 ($5.5\times10^{-5}$)   \\
X-class                       &  0.29 (0.10)                            &    0.40 ($2.1\times10^{-2}$)   &    0.51 ($2.2\times10^{-3}$)    \\
M-class                      &  0.31 (0.12)                            &   0.70 ($3.8\times10^{-5}$)    &    0.60 ($9.8\times10^{-4}$)    \\
Disk-center                &  0.40 ($3.3\times10^{-2}$)  &   0.70 ($1.2\times10^{-5}$)    &    0.78 ($8.8\times10^{-8}$)    \\
Limb                           &  0.004 (0.98)                          &   0.14 (0.47)                              &    0.21 (0.27)                               \\
Disk-center X-class &  0.63 ($1.9\times10^{-2}$)  &    0.66 ($1.2\times10^{-2}$)   &   0.86 ($4.4\times10^{-5}$)      \\
Disk-center M-class &  0.25 (0.36)                           &   0.69 ($2.2\times10^{-3}$)    &   0.58 ($1.6\times10^{-2}$)      \\
Limb X-class             &  -0.10 (0.69)                          &   0.02 (0.93)                              &   0.15 (0.56)                                \\
Limb M-class            &   0.41 (0.25)                           &    0.69 ($2.4\times10^{-2}$)   &   0.61 (0.06)                                \\
  \hline
\end{tabular}
\end{table}

Thus, the level of correlation between the flux loss and the total flux in the canceling features is approximately as high as the correlation between the flux loss and the flare strength. Therefore, it can account for the correlation between the flux loss and the GOES peak X-ray flux we observe in general, and with the X-class flares at disk-center locations in particular. However, we find no correlation between the flux loss and the total flux in the canceling features (or with the number of canceling features) for the limb X-class flares; thus it does not explain the correlation between the flux loss and the peak X-ray flux in this case. The correlation appears to work better for regions at disk-center locations than far from the disk center, suggesting that it is the radial flux loss that correlates best with the peak X-ray flux. We need vector magnetic field data to know the radial flux (not the nearly horizontal field seen in the line-of-sight magnetograms near the limb) to confirm this conjecture.

\subsection{Correlation between the Flux at the PIL and the X-Ray Flux}
\inlinecite{Schrijver07} measured the total magnetic flux ($R$) near the PIL and proposed this quantity as a measure for major flare potential rather than the total flux in an active region. In this work, the total flux in the PIL features can be considered as an equivalent to the parameter $R$. Figure 7 shows the relationship between the total magnetic flux in the PIL features against the total flux in the active region and the GOES peak X-ray flux. The total magnetic flux in the PIL features correlates with the total flux in the active region with the correlation coefficient of $\approx$ 0.60, which is in agreement with \inlinecite{Schrijver07}. Figure 7(b) indicates that large flares are associated with strong PIL for the flares at disk-center locations, consistent with the finding of \inlinecite{Schrijver07}.     

Such strong-gradient PILs are generally associated with flux emergence and flux cancellation driven by photospheric flows. The net flux change in the PIL features during its time evolution from before to after the flare shows that the PIL flux decreases before the flare outnumber the increases in X-class flares, while M-class flares show an approximately equal number of flares with PIL flux increases and decreases. This could hint that there is more canceling flux than emerging flux on PILs before the flare. However, after the flare we find approximately equal numbers of PIL flux decreases and increases in both X- and M-class flares. 

%
\begin{figure}[h]
\centering
\includegraphics[width=1.0\linewidth]{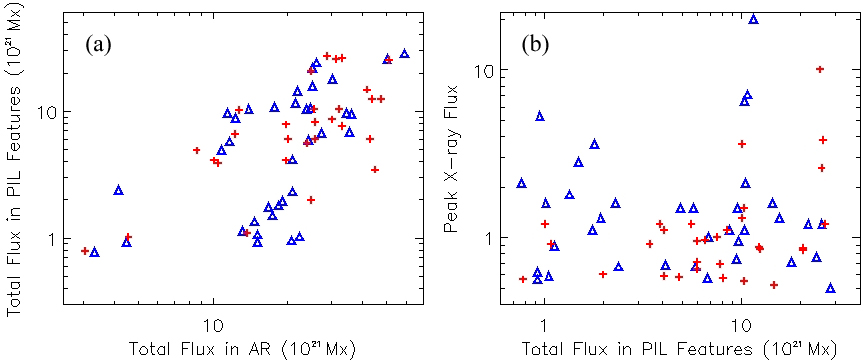}
\caption{(a) Total magnetic flux in the PIL features against the total flux in the active regions near the disk center (red pluses) and near the limb (blue triangles). (b) Total magnetic flux in the PIL features against the GOES peak X-ray flux. The peak X-ray flux of 1 corresponds to an X1-class flare.} 
\end{figure}

\subsection{Correlation between the Flux Loss and the Radial Position on the Solar Disk}
Figure 8 shows, as a function of the radial position of the region on the solar disk, the amount of magnetic flux loss, the total flux in the active region, the number of canceling features, and the total flux in the canceling features. This figure suggests that the magnetic flux loss, the total flux in the active region, and total flux in the canceling features are not correlated with the radial position of the region on the disk. The magnetic fields related to these categories have no apparent directional preferences, whereas Figures 4(c) and 4(d) show correlation between the flux changes and the radial position, consistent with nearly horizontal fields. On the other hand, Figure 8(c) does show correlation, indicating that the number of canceling features per active region is affected by foreshortening, decreasing towards the limb. Vector magnetic field data are needed to clarify whether the spatial resolution in the observations is the reason for this tendency.

%
\begin{figure}
\centering
\includegraphics[width=1.0\linewidth]{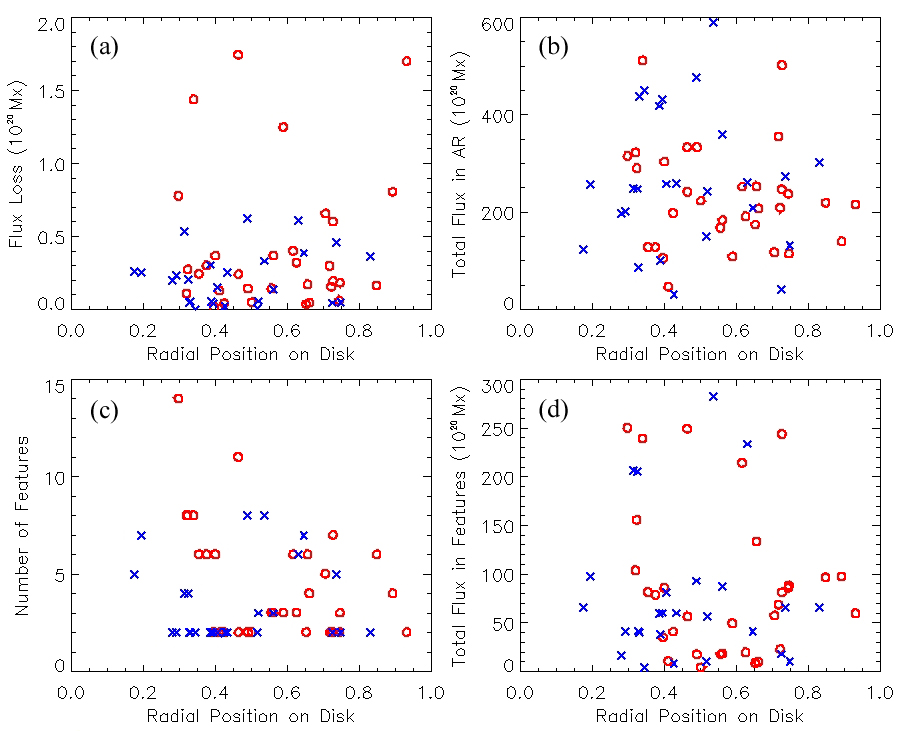}
\caption{(a) The amount of magnetic flux loss, (b) the total flux in the active region, (c) the number of canceling features, and (d) the total flux in the canceling features against the radial position of the region on the solar disk for M- (blue crosses) and X-class (red circles) flares.} 
\end{figure}

The Pearson correlation coefficients, summarized in Table 12, show weak anti-correlation between the total magnetic flux in the active region and its radial position. The number of canceling features indicates slightly higher, weak-to-moderate anti-correlation with the radial position of the region for the X-class flares only. The flux loss and the total flux in the canceling features show no statistically significant correlation with the radial position of the region. 

\begin{table}[bh!]
\caption{Pearson's correlation coefficients for the amount of magnetic flux loss, the total flux in the active region, the number of canceling features, and the total flux in the canceling features against the radial position of the region on the solar disk for M- and X-class flares. The $P$-values are given in parentheses.}
\label{T-simple}
\begin{tabular}{@{} l *{4}{c} @{}}     
  \hline                   
                                   &                Flux loss                    &              Total flux in               &              Number of                &        Total flux in           \\    
                                   &                                                    &            active region              &      canceling features          &  canceling features     \\                                     
  \hline
All                               &  0.18  (0.19)                             & -0.23 ($4.0\times10^{-2}$) & -0.18  (0.19)                           &  -0.08  (0.53)                \\
X-class                      &  0.10  (0.61)                             & -0.24 (0.14)                           & -0.43  ($4.8\times10^{-2}$)  &  -0.18  (0.33)                \\
M-class                      &  0.17 (0.40)                              & -0.21 (0.22)                           &  0.05   (0.80)                           & -0.02  (0.93)                 \\
  \hline
\end{tabular}
\end{table}

Cancellation events were not observed in all 77 flares. The absence of flux cancellation events generally corresponds to the active regions with flares far from the disk center; here the longitudinal component of the magnetic field is believed to be reduced in comparison to the disk-center locations because the photospheric field is approximately radial. However, there are exceptional cases where the disk-center flares do not involve flux cancellation or the near-limb flares show significant flux loss.  

The M5.0 flare on 17 January 2004 that occurred at $\approx$ 0.37 solar radii in NOAA 10540 did not involve flux cancellation events. The active region was diffuse with the total flux in all magnetic features lower than the median value of the total flux of all active regions analyzed. The same active region hosted another flare in our analysis, M6.1 on 20 January 2004, which also occurred at a near disk-center location ($\approx$ 0.32 solar radii) and showed a very small amount of flux loss with no multi-step flux cancellations. The active region was located at the coordinates of NOAA 10486 three rotations earlier, which was associated with Halloween's flare. After two following rotations (NOAA 10508 and 10523), the active region almost completely decayed. In another rotation, new flux emerged on the front side of the region and evolved into NOAA 10540, hosting a few C-class and M-class flares.

Three other flares (an M9.1 flare on 8 November 2001, an M9.1 flare on 22 July 2004, and an M6.0 flare on 5 September 2001) that took place at near disk-center locations ($\approx$ 0.48, 0.35, and 0.49 solar radii) did not involve multi-step flux cancellations. In the first flare, the active region had a diffuse structure and a short PIL. In the last two flares, the region was concentrated and had large total magnetic flux, but the total flux in the PIL features was lower than the median value of that for all the active regions analyzed.      

All active regions that hosted X-class flares in the near disk-center locations involved flux cancellation events. The absence of flux cancellation events around the X-class flares is generally seen in active regions with flares far from the disk center. However, we note the opposite cases in which some active regions with flares close to the limb locations involved significant flux cancellations during M- and X-class flares (see Tables 7 and 8 for details). They are the X20.0 flare on 2 April 2001 (NOAA 9393), M5.7 on 7 November 2001 (NOAA 9690), M7.1 on 26 December 2001 (NOAA 9742), M6.8 on 20 December 2002 (NOAA 10226), M7.6 on 26 October 2003 (NOAA 10484), X1.5 on 17 March 2003 (NOAA 10314), and X6.5 on 6 December 2006 (NOAA 10930). All of the active regions had concentrated, essentially balanced flux of both polarities and long and tight (strong) PILs.

\subsection{Flux Cancellations Versus Stepwise Flux Changes}
Do the flux cancellation events correlate with stepwise flux changes during the flares? The estimated amount of flux cancellation during some flares indicates a very large amount of flux loss (see Figure 5). We select the set of canceling features that contributed to the flux loss over the entire 4-h interval around a flare and consider also the subset of these features whose flux loss occurred only over the duration of the flare. Among these two sets of canceling features, we have counted the number of features which involve stepwise flux changes. We have compared 45 flares where at least one magnetic feature with stepwise flux change was found. Table 13 shows that only a small portion of magnetic features involving flux cancellation showed stepwise flux changes; thus we find no relationship between these two types of events. The lack of relationship between the stepwise field changes and the cancellation suggests that the photospheric cancellation process is independent of the abrupt flare-related field changes.

\begin{table}[h!]
\caption{The number of magnetic features, with and without stepwise flux changes, contributing to the flux cancellations over the entire 4-h interval around a flare and strictly over the duration of the flare.}
\label{T-simple}
\begin{tabular}{@{} l @{} *{4}{c} @{}}     
  \hline                   
Number                                &  \multicolumn{2}{c} {M-class flares}  &  \multicolumn{2}{c} {X-class flares}       \\
of features                            &    During flare     &         Total              &    During flare       &            Total              \\
  \hline
Stepwise flux change        &               0             &                2              &               4               &               13                \\
No stepwise flux change   &             19             &             47              &             37               &               79                \\
  \hline
\end{tabular}
\end{table}

\section{Discussion}

We have analyzed magnetic flux changes and studied flux cancellations related to 77 major flares. The results of this work can be summarized as follows.

We found that significantly more magnetic flux decreases occur during flares than increases. This result is consistent with the model of magnetic implosion and collapsing loop structure proposed by \inlinecite{Hudson08}. The change in the longitudinal magnetic field is larger in a region close to the limb, indicating that the involved field component is nearly horizontal. The X-class flares show larger flux changes and stronger bias towards flux decreases in comparison with M-class flares. 

Both net and total flux changes show an increasing trend with the flare strength. We find that a linear correlation between the flux changes and the GOES peak X-ray flux is dominated by X-class flares at limb locations, in agreement with the study of \inlinecite{Petrie10}. 

We identified episodes of flux cancellation and estimated the amount of flux canceled during three intervals around each flare: before the flare onset, during the flare, and after the flare end. We found that the amount of canceled flux is generally greater over the time during the flare than before and after the flare. This tendency is mostly contributed from X-class flares. Some of the strongest X-class flares showed significantly greater flux loss during the flare time. M-class flares did not show a significant bias in the amount of flux canceled during either time frame. 

The flux loss is approximately equally correlated with the number of canceling features and the total flux in these features. This leads us to conclude that the field strength and complexity are approximately equally important in the flux cancellation processes during flaring activity. The increasing trend of the flux losses as a function of the flare power is mostly explained by this correlation for the disk-center X-class flares, but not for the limb X-class flares. A less significant correlation between the flux losses and the total flux in the active region is also found for the flares at disk-center locations, dominated by the X-class flares. 

The flux loss is correlated with the peak X-ray flux for both disk-center and limb X-class flares. Thus, the correlation between the flare strength and both the flux losses and the flux changes is dominated by X-class flares.

We find that strong magnetic fields at PILs are associated with the occurrence of large flares, in agreement with \inlinecite{Schrijver07}. The correlation with the flare strength is higher for the flux loss than for the total flux in canceling features.

We did find contraction of the field towards the neutral line in the form of flux cancellation events, but we found no correlation between the flux cancellation and the abrupt, stepwise photospheric field changes associated with coronal field implosion during flares. We speculate that the stepwise changes and flux cancellation could be unrelated in general, as the stepwise flux changes may be caused by coronal magnetic restructuring while the flux cancellation may be caused by photospheric dynamics.

For future study, vector magnetic field data are needed to better interpret the longitudinal field changes. High-cadence, high-sensitivity full vector data from the Synoptic Optical Long-term Investigations of the Sun (SOLIS) and the Helioseismic and Magnetic Imager (HMI) will be very useful. Also a combined analysis of chromospheric and coronal data is needed to clarify the following: Are the canceling features part of the emerging U-loops or submerging $\Omega$-loops? Are the cancellations associated with reconnection? Does the reconnection occur above or below the photosphere?

\section*{Acknowledgments}

We thank Brian T. Welsch for providing the YAFTA feature-tracking code, fruitful discussions on this work, and useful comments. This work utilizes data obtained by the Global Oscillation Network Group (GONG) Program, managed by the National Solar Observatory, which is operated by AURA, Inc. under a cooperative agreement with the National Science Foundation. The data were acquired by instruments operated by the Big Bear Solar Observatory, High Altitude Observatory, Learmonth Solar Observatory, Udaipur Solar Observatory, Instituto de Astrof\'{\i}sica de Canarias, and Cerro Tololo Inter-American Observatory.

\end{article}		  
\end{document}